\documentclass[prl,twocolumn,superscriptaddress,aps]{revtex4-1}
\usepackage{amssymb,graphics,graphicx}
\usepackage{latexsym}
\usepackage{amsfonts}
\usepackage{amsmath}
\usepackage{float}

\newcommand{\be}{\begin{equation}}
\newcommand{\ee}{\end{equation}}
\newcommand{\bea}{\begin{eqnarray}}
\newcommand{\eea}{\end{eqnarray}}
\newcommand{\bma}{\begin{matrix}}
\newcommand{\ema}{\end{matrix}}

\newcommand{\bml}{\begin{mathletters}}
\newcommand{\eml}{\end{mathletters}}
\newcommand{\bes}{\begin{subequations}}
\newcommand{\ees}{\end{subequations}}

\newcommand{\bi}{\begin{itemize}}
\newcommand{\ei}{\end{itemize}}

\begin{document}
\title{A Possible Reason for $M_H \simeq 126$ GeV}
\author{Paul H. Frampton}
\email{paul.h.frampton@gmail.com}
\affiliation{Department of Physics and Astronomy, 
University of North Carolina, Chapel Hill, NC 27599-3255, USA}
\author{Pham Q. Hung}
\email{pqh@virginia.edu}
\affiliation{Department of Physics, University of Virginia,
Charlottesville, VA 22904-4714, USA;\\
Center for Theoretical and Computational Physics, 
Hue University College of Education, Hue, Vietnam}

\date{\today}

\begin{abstract}
It is speculated that a possible reason for the scalar
mass $M_H \simeq 126$ GeV is equality of the lifetimes
for vacuum decay and instanton-induced proton decay.
\end{abstract}

\maketitle

\section{Introduction}

In the standard model (SM) of particle theory 
there exists what has been called "the scandal of the fermion
masses" meaning that none of the twelve quark and lepton
masses have been satisfactorily explained by theory.
On the other hand, the masses of the spin-one bosons,
the photon, gluon, $W^{\pm}$ and $Z$ are well
understood in terms of symmetry breaking. Now, there
is one additional boson, with spin-zero, and the issue
is whether its mass can be explained.

\bigskip

\noindent
Surely the most dramatic discovery \cite{ATLAS,CMS} 
in experimental high-energy
physics so far this century is that of the Brout-Englert-Higgs
\cite{Englert,Higgs1,Higgs2,Guralnik,Higgs3} 
boson (H) which completes the discovery of all the particles in the
minimal standard model. Equally dramatic is the non-discovery
of any other particle in the mass range which has so far been
investigated. Given what we now know, it is very reasonable
to assume there are no other particles at low energies, and
that all that exists is the $SU(3)_C \times SU(2)_L \times
U(1)_Y$ standard model with three quark-lepton
families and only one scalar whose mass is measured as
$M_H \simeq 126$ GeV.

\bigskip

\noindent
At large scales, classical gravity is well described by general
relativity and we assume that this is the theory which governs
the behavior of the universe as a whole. Although large black
holes are known to exist, we will ignore the effects of possible
virtual black holes in microscopic processes because there
is no fully satisfactory theory of quantum gravity at small scales.

\bigskip

\noindent
In the present note, we shall offer a speculation which
provides a possible reason for $M_H \simeq 126$ GeV, as well as
saying something about the fate of dark energy in the
extreme future of the Universe.

\bigskip

\noindent
In gauge field theory, an important formal development was
the discovery of the instanton solution \cite{Belavin}. For
example, this was once proposed \cite{CallanDashenGross}
as the key to understanding the QCD of strong interactions.
In this note, the instantons of electroweak interactions
will play a central role.

\section{Lifetime $\tau_{EW}$ of Metastable Vacuum}
\noindent

The decay of the metastable electroweak vacuum into 
the true ground state was first studied
as a function of $M_H$ in 
\cite{Kobzarev,Kobzarev2,
Linde,Weinberg,Frampton,Frampton2,Coleman,CallanColeman}, 
including
only boson loops. Another kind of instability was
first emphasized in \cite{krasnikov,Hung,PolitzerWolfram,cabibbo} where the 
presence of heavy fermion loops could make the effective potential 
unbounded from below. The conditions for vacuum stability were 
further improved by summing over leading logs in the effective 
potential \cite{sher1,sher11,sher2,sher3,sher4}. 
It was also emphasized in various subsequent works such 
as e.g. \cite{isidori} that the so-called unbounded potential is 
only an artifact since near the energy scale where instability 
is supposed to occur the top quark Yukawa coupling becomes 
small \cite{sher2} and/or new physics enters (see e.g. \cite{sher5}) and the potential acquires 
a true vacuum at a value of the field much larger than the 
electroweak scale.

\bigskip

\noindent
The discovery of the H scalar with mass $M_H \simeq 126$ GeV has
precipitated a careful reevaluation of the 
electroweak vacuum stability
and the result is especially 
interesting, that the
vacuum is not perfectly stable. The result depends mainly on the
mass of the top quark $M_t$ and on $M_H$. 
This metastability of the vacuum depends sensitively
on the values of these parameters. With the observed values
$M_t = 173.36 \pm 0.65 \pm 0.3$ GeV and 
$M_H = 125.66 \pm 0.34$ GeV,
the vacuum is metastable with an extremely long lifetime discussed
below. If we keep one of the two fixed and vary the other,
 the sensitivity
from \cite{BDGGSSS} is seen to be that the vacuum would become
absolutely stable if $M_t$ were reduced to $171$ GeV, 
or if $M_H$ were increased to $130$ GeV.

\bigskip

This proximity to criticality is sufficiently striking that
it is likely informing us profoundly about theoretical physics.
It can be speculated that there is a statistical
basis provided by a multiverse interpretation but such an idea
seems untestable. Here we aim to exploit only facts which are on
a firm basis.

\bigskip

In \cite{BDGGSSS} two different cosmological futures 
are considered:
the $\Lambda$CDM model and a $CDM$ model where the dark energy
disappears. For these cases the lifetime of the electroweak
vacuum for the observed values of $M_H$ and $M_t$ is
respectively
\begin{equation}
10^{400} < \tau_{EW} < 10^{700}~~~ {\rm years} 
\label{LCDM}
\end{equation}
for the $\Lambda$CDM model, and
\begin{equation}
10^{100} < \tau_{EW} < 10^{400}~~~ {\rm years} 
\label{CDM}
\end{equation}
for the CDM model.

\bigskip

\section{Lifetime $\tau_p$ of Proton}

According to the results of the Super-Kamiokande experiment
\cite{SuperK}, the lower bound on some proton decay
partial lifetimes are
$\tau (p \rightarrow e^+ \pi^0) \geq 8.2 \times 10^{33}$ years
and 
$\tau (p \rightarrow \mu^+ \pi^0) \geq 6.6 \times 10^{33}$ years.
In grand unified theories, {\it e.g.} \cite{GG}, 
the lifetime was initially predicted as 
$\tau (p \rightarrow e^+ \pi^0) \sim 10^{31}$ years
which disagrees with experiment. We 
shall assume there is no grand unification which
induces proton decay.
As stated in the Introduction, we shall neglect also the effects
of virtual black holes which could \cite{Sakharov} otherwise
give $\tau_p \sim 10^{50}$ years.

\bigskip

\noindent
Perturbatively, the standard model conserves baryon number (B) and
lepton number (L) but, as first noted by 't Hooft \cite{Hooft,Hooft2},
non-perturbative
instantons induce violations of B and L while preserving $(B-L)$.
This leads to instanton-induced proton decay. The 
rate is suppressed by a factor
\begin{equation} 
exp(- 4\pi/ \alpha_2) = exp(- 4\pi \sin^2 \Theta_W / \alpha_{em})
\label{suppression}
\end{equation}

\bigskip

\noindent
The lifetime for proton decay $\tau_p$ is proportional
to the inverse of this rate and so
\begin{eqnarray}
\tau_p  &\propto& exp (+ 4\pi \sin^2 \Theta_W / \alpha_{em}) \nonumber  \\
& \simeq &  exp (+ 371) \simeq 10^{+160} 
\label{taup}
\end{eqnarray}

\noindent
The determinant of fermion zero modes in \cite{Hooft2} will
affect only the prefactor so that to a zeroth approximation
\begin{equation}
\tau_p \sim 10^{160} ~~~ {\rm years}
\label{taup2}
\end{equation}
which is such a long lifetime that the difference from
absolute stability may {\it prima facie} seem only academic. 
Nevertheless, we shall argue that the finite lifetime 
in Eq.(\ref{taup2}) is very important.
A word of caution is in order here.
Note that even if the prefactor were calculated accurately for 
the amplitude of the proton decay, it would not so drastically 
change the estimate Eq.(\ref{taup2}) to bring it outside the 
range permitted by Eq. (\ref{CDM}). It is thus very
reasonable to expect that 
the instanton-induced proton decay lifetime lies comfortably 
within the CDM range. As a result, conservatively, we expect

\begin{equation}
10 ^{100} < \tau_p > 10^{400} {\rm years}
\end{equation}

(By enlarging the SM gauge group, it might be possible to shorten
considerably the instanton-induced proton lifetime \cite{wagner}. 
However, we limit ourselves strictly
to the minimal SM as we have stressed in the second paragraph of this paper.)

\section{A possible reason}

Both $\tau_{EW}$ and $\tau_p$ are related to electroweak
instantons, and both represent major decay processes in 
the extreme future of the Universe. Based on the above
observations, we make the following conjecture
\begin{equation}
\tau_{EW} (M_H) = \tau_p \, ,
\label{tautau}
\end{equation}
where $\tau_{EW}$ has a sensitive dependence 
on the Higgs mass $M_H$.
Since $\tau_p$ is essentially fixed by the $SU(2)_L$ gauge 
coupling by Eq.(\ref{taup}), the equality 
Eq.(\ref{tautau}) amounts to a 
determination of $M_H$ in terms of the proton lifetime.
Comparison of $\tau_{EW}$ (CDM) with Eq.(\ref{taup2})
then shows that the value $M_H \simeq 126$ GeV is required
for this equality. If $M_H$ were increased to
{\it e.g.} $130$ GeV, $\tau_{EW}$ would become infinite. If $M_H$ were  
reduced to {\it e.g.} $122$ GeV, one would find $\tau_{EW} \ll \tau_p$.
The fact that Eq.(\ref{tautau}) requires $\tau_{EW}$ (CDM)
rather that $\tau_{EW}$ ($\Lambda$CDM) from \cite{BDGGSSS}
suggests that the
dark energy will disappear before cosmic time $t = 10^{160}$
years, although a more careful analysis of Eq. (\ref{tautau})
could sharpen this prediction about the extreme future of 
the Universe.

Needless to say, the deep reason behind the conjecture
(\ref{tautau}) is beyond our grasp at the present time
and it would be very interesting if one could be found
in the future.

\bigskip

\noindent
As alluded to above, the true minimum is located 
at a value of the Higgs field orders of magnitude larger  
than for our metastable vacuum which is located at the electroweak scale. 
In the stable vacuum, quarks and leptons (and $W^{\pm}$, $Z$)
will be supermassive and bound states very small compared to
in the known universe, leading inevitably
to a preponderance of black holes.
On the other hand, if the protons have decayed long
before the phase transition
to the true ground state, as in the $\Lambda$CDM model, there 
will be no remaining quarks,
only leptons $e^{\pm}$, $\nu$ and photons $\gamma$ and, assuming
the dark matter does not change, far less production of black holes. 
Such might not be the case if the proton decays when the phase 
transition to the true ground state occurs.
This type of discussion may underwrite a rationale for the
conjectured equality in Eq.(\ref{tautau}).

\bigskip

\noindent
The true ground state is stabilized by an incompletely 
understood mechanism and, 
as we have argued for the CDM case in a far distant future, 
it has vanishing vacuum energy since the dark energy has  
disappeared. Once gravity is included, and therefore absolute
potential energy acquires significance,
the decay is from the present vacuum which 
is dominated by dark energy with a relatively very tiny energy 
density to one with vanishing energy density, and it is tempting then
to speculate even further that the mechanism that stabilizes the true vacuum 
also sets the energy gap between the two vacua equal to  
the dark energy density.

\section{Discussion}

Our Eq.(\ref{tautau}) equates the two longest time scales
associated with the standard model. It has the advantage
of using only facts which are known. We could 
hold $M_H$ fixed, and vary $M_t$,
but $M_H$ is the more fundamental because the scalar is 
an exceptional particle in the standard model,
being the only one with zero spin.
If our present discussion is correct, 
it implies that the boson masses, rather than the fermion masses,
are the more tractable.

\bigskip

\section{Acknowledgments}
PHF was supported in part by US DOE grant DE-FG02-06ER41418.
PQH was supported in part by US DOE grant DE-FG02-97ER41027.


\begin{thebibliography}{99}
\bibitem{ATLAS}
G. Aad, {\it et al.} (ATLAS Collaboration),
Phys. Lett. {\bf B716,} 1 (2012).
{\tt arXiv:1207.7214 [hep-ex]}.
\bibitem{CMS}
S. Chatrchyan, {\it et al.} (CMS Collaboration),
Phys. Lett. {\bf B716,} 30 (2012).
{\tt arXiv:1207.7235 [hep-ex]}.
\bibitem{Englert}
F. Englert and R. Brout, Phys. Rev. Lett. {\bf 13,} 321 (1964).
\bibitem{Higgs1}
P.W. Higgs, Phys. Lett. {\bf 12,} 132 (1964).
\bibitem{Higgs2}
P.W. Higgs, Phys. Rev. Lett. {\bf 13,} 508 (1964). 
\bibitem{Guralnik}
G.S. Guralnik, C.R. Hagen and T.W.B. Kibble,
Phys. Rev. Lett. {\bf 13,} 585 (1964).
\bibitem{Higgs3}
P.W. Higgs, Phys. Rev. {\bf 145,} 1156 (1966).
\bibitem{Belavin}
A.A. Belavin, A.M. Polyakov, A.S. Schwartz, and Yu.S. Tyupkin,
Phys. Lett. {\bf B59,} 85 (1975).
\bibitem{CallanDashenGross}
C.G. Callan Jr., R.F. Dashen, and D.J. Gross, 
Phys. Rev. {\bf D17,} 2717 (1977).
\bibitem{Kobzarev}
I. Yu. Kobzarev, L.B. Okun, and M.B. Voloshin, 
Yad. Fiz. {\bf 20,} 1229 (1974). 
\bibitem{Kobzarev2}
I. Yu. Kobzarev, L.B. Okun, and M.B. Voloshin, 
Sov. J. Nucl. Phys. {\bf 20,} 644 (1975).
\bibitem{Linde}
A.D. Linde, JETP Lett. {\bf 23,} 64 (1976).
\bibitem{Weinberg}
S. Weinberg, Phys. Rev. Lett. {\bf 36,} 294 (1976).
\bibitem{Frampton}
P.H. Frampton, Phys. Rev. Lett. {\bf 37,} 1378 (1976).
\bibitem{Frampton2}
P.H. Frampton, Phys. Rev. {\bf D15,} 2922 (1977).
\bibitem{Coleman}
S. Coleman, Phys. Rev. {\bf D15,} 2929 (1977).
\bibitem{CallanColeman}
C.G. Callan, Jr., and S. Coleman, 
Phys. Rev. {\bf D16,} 1762 (1977).
\bibitem{krasnikov}
N.~V.~Krasnikov,
  Yad.\ Fiz.\  {\bf 28}, 549 (1978).
\bibitem{Hung}
P.Q. Hung, Phys. Rev. Lett. {\bf 42,} 873 (1979)
\bibitem{PolitzerWolfram}
H.D. Politzer and S. Wolfram, Phys. Lett. {\bf B82,} 242 (1979).
\bibitem{cabibbo}
N.~Cabibbo, L.~Maiani, G.~Parisi and R.~Petronzio,
Nucl.\ Phys.\ B {\bf 158}, 295 (1979).
\bibitem{sher1}
R.A. Flores and M. Sher, Phys. Rev. {\bf D27,} 1679 (1983).
\bibitem{sher11}
M.~J.~Duncan, R.~Philippe and M.~Sher,
Phys.\ Lett.\ B {\bf 153}, 165 (1985)
[Erratum-ibid.\  {\bf 209B}, 543 (1988)].
\bibitem{sher2}
M. Sher, Phys. Reports {\bf 179,} 273 (1989).
\bibitem{sher3}
M. Lindner, M. Sher, and H.W. Zaglauer, Phys. Lett. {\bf 228,} 139 (1989).
\bibitem{sher4}
M. Sher, Phys. Lett. {\bf B317,} 159 (1993).
{\tt arXiv:hep-ph/9307342}.
\bibitem{isidori}
G. Isidori, G. Ridolfi, and A. Strumia, Nucl. Phys. {\bf B609,} 387 (2001).
{\tt arXiv:hep-ph/0104016}.
\bibitem{sher5}
P.Q. Hung and M. Sher, Phys. Lett. {\bf B374,} 138 (1996).
{\tt arXiv:hep-ph/9512313}.
\bibitem{BDGGSSS}
D. Buttazzo, G. Degrassi, P.P. Giardino, G.F. Giudice, F. Sala,
A. Salvio and A. Strumia. {\tt arXiv:1307.3536[hep-ph]}
\bibitem{SuperK}
H. Nishino, {\it et al.} (Super-Kamiokande Collaboration).
Phys. Rev. {\bf D85,} 112001 (2012).
{\tt arXiv:1203.4030 [hep-ex]}.
\bibitem{GG}
H. Georgi and S.L. Glashow, Phys. Rev. Lett. {\bf 32,} 438 (1974).
\bibitem{Sakharov}
A.D Sakharov, Pisma Zh. Eksp. Teor. Fiz. {\bf 5,} 32 (1967);
JETP Lett. {\bf 5,} 24 (1967).
\bibitem{Hooft}
G. 't Hooft, Phys. Rev. Lett. {\bf 373,} 8 (1976).
\bibitem{Hooft2}
G. 't Hooft, Phys. Rev. {\bf D14,} 3432 (1976).
\bibitem{wagner}
D.~E.~Morrissey, T.~M.~P.~Tait and C.~E.~M.~Wagner,
Phys.\ Rev.\ D {\bf 72}, 095003 (2005)
{\tt arXiv:hep-ph/0508123}.
\end{thebibliography}
\end{document}